\RequirePackage{lineno}

\documentclass[aps,prd,twocolumn,superscriptaddress,showpacs,floatfix,nofootinbib]{revtex4}

\usepackage{comment}
\usepackage{graphicx}
\usepackage[dvips]{color} 
\usepackage{bm}

\newcommand{\MET}{ $/\!\!\!\!\!E_{T}$}

\begin{document}

\vspace{2cm}

\title{Search for Standard Model Higgs Boson Production in Association
 with a $W$ Boson using a Neural Network Discriminant at CDF}

 \affiliation{Institute of Physics, Academia Sinica, Taipei, Taiwan 11529, Republic of China} 
\affiliation{Argonne National Laboratory, Argonne, Illinois 60439} 
\affiliation{University of Athens, 157 71 Athens, Greece} 
\affiliation{Institut de Fisica d'Altes Energies, Universitat Autonoma de Barcelona, E-08193, Bellaterra (Barcelona), Spain} 
\affiliation{Baylor University, Waco, Texas  76798} 
\affiliation{Istituto Nazionale di Fisica Nucleare Bologna, $^y$University of Bologna, I-40127 Bologna, Italy} 
\affiliation{Brandeis University, Waltham, Massachusetts 02254} 
\affiliation{University of California, Davis, Davis, California  95616} 
\affiliation{University of California, Los Angeles, Los Angeles, California  90024} 
\affiliation{University of California, San Diego, La Jolla, California  92093} 
\affiliation{University of California, Santa Barbara, Santa Barbara, California 93106} 
\affiliation{Instituto de Fisica de Cantabria, CSIC-University of Cantabria, 39005 Santander, Spain} 
\affiliation{Carnegie Mellon University, Pittsburgh, PA  15213} 
\affiliation{Enrico Fermi Institute, University of Chicago, Chicago, Illinois 60637}
\affiliation{Comenius University, 842 48 Bratislava, Slovakia; Institute of Experimental Physics, 040 01 Kosice, Slovakia} 
\affiliation{Joint Institute for Nuclear Research, RU-141980 Dubna, Russia} 
\affiliation{Duke University, Durham, North Carolina  27708} 
\affiliation{Fermi National Accelerator Laboratory, Batavia, Illinois 60510} 
\affiliation{University of Florida, Gainesville, Florida  32611} 
\affiliation{Laboratori Nazionali di Frascati, Istituto Nazionale di Fisica Nucleare, I-00044 Frascati, Italy} 
\affiliation{University of Geneva, CH-1211 Geneva 4, Switzerland} 
\affiliation{Glasgow University, Glasgow G12 8QQ, United Kingdom} 
\affiliation{Harvard University, Cambridge, Massachusetts 02138} 
\affiliation{Division of High Energy Physics, Department of Physics, University of Helsinki and Helsinki Institute of Physics, FIN-00014, Helsinki, Finland} 
\affiliation{University of Illinois, Urbana, Illinois 61801} 
\affiliation{The Johns Hopkins University, Baltimore, Maryland 21218} 
\affiliation{Institut f\"{u}r Experimentelle Kernphysik, Universit\"{a}t Karlsruhe, 76128 Karlsruhe, Germany} 
\affiliation{Center for High Energy Physics: Kyungpook National University, Daegu 702-701, Korea; Seoul National University, Seoul 151-742, Korea; Sungkyunkwan University, Suwon 440-746, Korea; Korea Institute of Science and Technology Information, Daejeon, 305-806, Korea; Chonnam National University, Gwangju,500-757, Korea; Chonbuk National University, Jeonju 561-756, Korea} 
\affiliation{Ernest Orlando Lawrence Berkeley National Laboratory, Berkeley, California 94720} 
\affiliation{University of Liverpool, Liverpool L69 7ZE, United Kingdom} 
\affiliation{University College London, London WC1E 6BT, United Kingdom} 
\affiliation{Centro de Investigaciones Energeticas Medioambientales y Tecnologicas, E-28040 Madrid, Spain} 
\affiliation{Massachusetts Institute of Technology, Cambridge, Massachusetts  02139} 
\affiliation{Institute of Particle Physics: McGill University, Montr\'{e}al, Qu\'{e}bec, Canada H3A~2T8; Simon Fraser University, Burnaby, British Columbia, Canada V5A~1S6; University of Toronto, Toronto, Ontario, Canada M5S~1A7; and TRIUMF, Vancouver, British Columbia, Canada V6T~2A3} 
\affiliation{University of Michigan, Ann Arbor, Michigan 48109} 
\affiliation{Michigan State University, East Lansing, Michigan  48824}
\affiliation{Institution for Theoretical and Experimental Physics, ITEP, Moscow 117259, Russia} 
\affiliation{University of New Mexico, Albuquerque, New Mexico 87131} 
\affiliation{Northwestern University, Evanston, Illinois  60208} 
\affiliation{The Ohio State University, Columbus, Ohio  43210} 
\affiliation{Okayama University, Okayama 700-8530, Japan} 
\affiliation{Osaka City University, Osaka 588, Japan} 
\affiliation{University of Oxford, Oxford OX1 3RH, United Kingdom} 
\affiliation{Istituto Nazionale di Fisica Nucleare, Sezione di Padova-Trento, $^z$University of Padova, I-35131 Padova, Italy} 
\affiliation{LPNHE, Universite Pierre et Marie Curie/IN2P3-CNRS, UMR7585, Paris, F-75252 France} 
\affiliation{University of Pennsylvania, Philadelphia, Pennsylvania 19104}
\affiliation{Istituto Nazionale di Fisica Nucleare Pisa, $^{aa}$University of Pisa, $^{bb}$University of Siena and $^{cc}$Scuola Normale Superiore, I-56127 Pisa, Italy} 
\affiliation{University of Pittsburgh, Pittsburgh, Pennsylvania 15260} 
\affiliation{Purdue University, West Lafayette, Indiana 47907} 
\affiliation{University of Rochester, Rochester, New York 14627} 
\affiliation{The Rockefeller University, New York, New York 10021} 
\affiliation{Istituto Nazionale di Fisica Nucleare, Sezione di Roma 1, $^{dd}$Sapienza Universit\`{a} di Roma, I-00185 Roma, Italy} 

\affiliation{Rutgers University, Piscataway, New Jersey 08855} 
\affiliation{Texas A\&M University, College Station, Texas 77843} 
\affiliation{Istituto Nazionale di Fisica Nucleare Trieste/Udine, I-34100 Trieste, $^{ee}$University of Trieste/Udine, I-33100 Udine, Italy} 
\affiliation{University of Tsukuba, Tsukuba, Ibaraki 305, Japan} 
\affiliation{Tufts University, Medford, Massachusetts 02155} 
\affiliation{Waseda University, Tokyo 169, Japan} 
\affiliation{Wayne State University, Detroit, Michigan  48201} 
\affiliation{University of Wisconsin, Madison, Wisconsin 53706} 
\affiliation{Yale University, New Haven, Connecticut 06520} 
\author{T.~Aaltonen}
\affiliation{Division of High Energy Physics, Department of Physics, University of Helsinki and Helsinki Institute of Physics, FIN-00014, Helsinki, Finland}
\author{J.~Adelman}
\affiliation{Enrico Fermi Institute, University of Chicago, Chicago, Illinois 60637}
\author{T.~Akimoto}
\affiliation{University of Tsukuba, Tsukuba, Ibaraki 305, Japan}
\author{B.~\'{A}lvarez~Gonz\'{a}lez$^t$}
\affiliation{Instituto de Fisica de Cantabria, CSIC-University of Cantabria, 39005 Santander, Spain}
\author{S.~Amerio$^z$}
\affiliation{Istituto Nazionale di Fisica Nucleare, Sezione di Padova-Trento, $^z$University of Padova, I-35131 Padova, Italy} 

\author{D.~Amidei}
\affiliation{University of Michigan, Ann Arbor, Michigan 48109}
\author{A.~Anastassov}
\affiliation{Northwestern University, Evanston, Illinois  60208}
\author{A.~Annovi}
\affiliation{Laboratori Nazionali di Frascati, Istituto Nazionale di Fisica Nucleare, I-00044 Frascati, Italy}
\author{J.~Antos}
\affiliation{Comenius University, 842 48 Bratislava, Slovakia; Institute of Experimental Physics, 040 01 Kosice, Slovakia}
\author{G.~Apollinari}
\affiliation{Fermi National Accelerator Laboratory, Batavia, Illinois 60510}
% ...

\collaboration{CDF Collaboration\footnote{With visitors from $^a$University of Massachusetts Amherst, Amherst, Massachusetts 01003,
$^b$Universiteit Antwerpen, B-2610 Antwerp, Belgium, 
$^c$University of Bristol, Bristol BS8 1TL, United Kingdom,
$^d$Chinese Academy of Sciences, Beijing 100864, China, 
$^e$Istituto Nazionale di Fisica Nucleare, Sezione di Cagliari, 09042 Monserrato (Cagliari), Italy,
$^f$University of California Irvine, Irvine, CA  92697, 
$^g$University of California Santa Cruz, Santa Cruz, CA  95064, 
$^h$Cornell University, Ithaca, NY  14853, 
$^i$University of Cyprus, Nicosia CY-1678, Cyprus, 
$^j$University College Dublin, Dublin 4, Ireland,
$^k$University of Edinburgh, Edinburgh EH9 3JZ, United Kingdom, 
$^l$University of Fukui, Fukui City, Fukui Prefecture, Japan 910-0017
$^m$Kinki University, Higashi-Osaka City, Japan 577-8502
$^n$Universidad Iberoamericana, Mexico D.F., Mexico,
$^o$University of Iowa, Iowa City, IA  52242,
$^p$Queen Mary, University of London, London, E1 4NS, England,
$^q$University of Manchester, Manchester M13 9PL, England, 
$^r$Nagasaki Institute of Applied Science, Nagasaki, Japan, 
$^s$University of Notre Dame, Notre Dame, IN 46556,
$^t$University de Oviedo, E-33007 Oviedo, Spain, 
$^u$Texas Tech University, Lubbock, TX  79609, 
$^v$IFIC(CSIC-Universitat de Valencia), 46071 Valencia, Spain,
$^w$University of Virginia, Charlottesville, VA  22904,
$^x$Bergische Universit\"at Wuppertal, 42097 Wuppertal, Germany,
$^{ff}$On leave from J.~Stefan Institute, Ljubljana, Slovenia, 
}}
\noaffiliation


\date{\today}

\begin{abstract}
We present a search for standard model Higgs boson production in
association with a $W$ boson in proton-antiproton collisions
($p\bar{p}\rightarrow W^\pm H \rightarrow \ell\nu b\bar{b}$) at a
center of mass energy of 1.96 TeV. The search employs data collected
with the CDF II detector that correspond to an integrated luminosity
of approximately 1.9 fb$^{-1}$.  We select events consistent with a
signature of a single charged lepton ($e^\pm/\mu^\pm$), missing transverse
energy, and two jets. Jets corresponding to bottom quarks are
identified with a secondary vertex tagging method, a jet probability tagging method, and a neural network
filter. We use kinematic information in an artificial neural network to 
improve discrimination between signal and background compared to previous analyses. 
The observed number of events and the neural network output distributions are consistent with 
the standard model background
expectations, and we set 95\% confidence level upper limits on the
production cross section times branching fraction ranging from 1.2 to 1.1~pb or 
7.5 to 102 times the standard model expectation for
Higgs boson masses from 110 to $150\,\mathrm{GeV}/c^2$, respectively.
\end{abstract}


\maketitle

\section{Introduction}

Standard electroweak theory predicts a single fundamental scalar
particle, the Higgs boson, which arises as a result of spontaneous
electroweak symmetry breaking~\cite{Higgs:1964pj}. 
However, the Higgs boson is the only fundamental standard model particle which has not been directly observed by experiments.
The current experimental constraint on the Higgs boson mass, $m_H >
114.4\,\mathrm{GeV}/c^2$ at 95\% confidence level (C.L.), comes from
direct Higgs boson searches at LEP experiments~\cite{Barate:2003sz}.
Global fits to electroweak measurements 
exclude masses above
$154\,\mathrm{GeV}/c^2$ at 95\% confidence level~\cite{Alcaraz:2008mx}.

At the Tevatron $p\bar p$ collider at Fermilab, the
next-to-leading-order (NLO) Higgs boson production cross section
is about 10 times larger for gluon fusion than for $WH$ associated
production, and the cross section for $WH$ is about twice that of
$ZH$~\cite{Han:1991ia}.  The Higgs boson decay branching fraction is
dominated by $H\rightarrow b\bar{b}$ for $m_H<135\,\mathrm{GeV}/c^2$
and by $H\rightarrow W^+W^-$ for
$m_H>135\,\mathrm{GeV}/c^2$~\cite{Djouadi:1997yw}.  Background
$b\bar b$ production processes have QCD cross sections at least 4
orders of magnitude greater than that of Higgs boson
production~\cite{Abulencia:2006ps}, and this renders searches in the
$gg\rightarrow H
\rightarrow b\bar{b}$ channel nonviable. Requiring the
leptonic decay of the associated $W$ boson reduces the huge $b\bar b$
background rate. As a result, $WH\rightarrow \ell
\nu b\bar{b}$ is one of the most favorable channels for
a low mass Higgs boson search\footnote{In this paper, lepton
($\ell$) denotes electron ($e^\pm$) or muon ($\mu^\pm$), and neutrino
($\nu$) denotes electron neutrino ($\nu_e$) or muon neutrino
($\nu_\mu$) or their antiparticles.}, and it forms an important component of the combined search for the Higgs boson at the Tevatron.

Searches for $WH\rightarrow \ell\nu b\bar{b}$ at 
$\sqrt{s}=1.96\,\rm{TeV}$ have been most recently reported by CDF~\cite{Aaltonen:2007wx,Aaltonen:2008zx} 
and D0~\cite{new_d0} using data corresponding to an integrated luminosity of
955~pb$^{-1}$ and 440~pb$^{-1}$, respectively.  
In this paper, we present an update on the search for $WH\rightarrow
\ell\nu b\bar{b}$ production at CDF using about 1.9 fb$^{-1}$ of data
and improved analysis techniques. 
The $b$ hadrons which arise from $b\bar{b}$ quark pairs in this final state 
can be identified through special algorithms based on the tracking information 
for particles within hadronic jets, a procedure known as $b$-tagging. 
We have optimized the $b$-tagging algorithms to increase the acceptance for double $b$-tagged events.
In addition we increase the 
signal acceptance by including the electrons going into the forward 
region of the detector and introduce a
multivariate discriminant technique using a neural network (NN) to reduce large background 
contamination after the event selection.

The paper is organized as follows. 
Section~\ref{sec:detector} describes the CDF II
detector. The event selection criteria are explained in Sec.~\ref{sec:eventSelection}.
In Sec.~\ref{sec:btag} the $b$-tagging
algorithms {\sc secvtx}~\cite{Acosta:2004hw}, neural net $b$-tagging filter, and jet probability~\cite{Abulencia:2006kv} are discussed in
detail.  Contributions from the standard model (SM) background in the $WH\rightarrow \ell\nu b\bar{b}$ channel are
calculated in Sec.~\ref{sec:bkg} for various sources. In
Sec.~\ref{sec:Acceptance}, signal acceptance and systematic
uncertainties are estimated. The neural network discriminant technique is described in Sec.~\ref{sec:neuralnet}. 
The results and their statistical interpretation are presented in
Sec.~\ref{sec:results}. Finally, our conclusions are presented in Sec.~\ref{sec:conclusions}.

\section{CDF II Detector}
\label{sec:detector}
The CDF II detector~\cite{Acosta:2004yw} geometry is described using a 
cylindrical coordinate system.  The $z$-axis follows the
proton direction, and the polar angle $\theta$ is usually expressed
through the pseudorapidity $\eta = -\ln(\tan(\theta/2))$.  The
detector is approximately symmetric in $\eta$ and in the azimuthal
angle~$\phi$.  The transverse energy is defined as $E_T=E\sin \theta$,
and the transverse momentum $p_T=p \sin \theta$.

Charged particles are tracked by a system of silicon microstrip
detectors and a large open cell drift chamber in the region
$|\eta|\leq 2.0$ and $|\eta|\leq 1.0$, respectively.  The tracking
detectors are immersed in a $1.4\,\mathrm{T}$ solenoidal magnetic
field aligned coaxially with the incoming beams, allowing measurement
of charged particle momentum transverse to the beamline ($p_T$).

The transverse momentum resolution is measured to be $\delta p_T/p_T
\approx 0.07\% \cdot p_T$(GeV) for the combined tracking system.
The resolution on the track impact parameter ($d_0$), the distance from
the beamline axis to the track at the track's closest approach in the
transverse plane, is $\sigma(d_0) \approx 40\,\mu{\rm m}$, of which about
$30\,\mu{\rm m}$ is due to the transverse size of the
Tevatron beam itself.

Outside of the tracking systems and the solenoid, segmented
calorimeters with projective tower geometry are used to reconstruct
electromagnetic showers and hadronic
jets~\cite{Balka:1987ty,Bertolucci:1987zn,Albrow:2001jw} over the
pseudorapidity range $|\eta|<3.6$.  A transverse energy is measured 
in each calorimeter tower where the polar
angle ($\theta$) is calculated using the measured $z$ position of the event
vertex and the tower location.

Contiguous groups of calorimeter towers with signals are
identified and summed together into an energy cluster.  Electron
candidates are identified in the central electromagnetic calorimeter
(CEM) or in the forward, known as the plug, electromagnetic calorimeter (PEM) 
as isolated, mostly electromagnetic clusters that match a track
in the pseudorapidity range $|\eta|<1.1$ and $1.1<|\eta| < 2.0$, respectively.  The electron transverse
energy is reconstructed from the electromagnetic cluster with a
precision $\sigma(E_T)/E_T \approx 13.5\%/\sqrt{E_T(\mathrm{GeV})} \oplus
2\%$ for central electrons~\cite{Balka:1987ty} and $\sigma(E_T)/E_T = 16.0\%/\sqrt{E_T(\mathrm{GeV})} \oplus
2\%$ for plug electrons.  Jets are identified as a group of
electromagnetic (EM) and hadronic (HAD) calorimeter clusters which
fall within a cone of radius $\Delta{R} \approx\sqrt{(\Delta
\phi)^2 + (\Delta \eta)^2} \leq 0.4$ units around a high-$E_T$ seed
cluster~\cite{Abe:1991ui}.  Jet energies are corrected for calorimeter
nonlinearity, losses in the gaps between towers and
multiple primary interactions. The jet energy resolution is
approximately $\sigma(E_T) \approx \left[0.1 E_T + 1.0~GeV\right]$~\cite{Bhatti:2005ai}.

Muon candidates are detected in three separate subdetectors.
After at least five interaction lengths in the calorimeter, the large angle muons first
encounter four layers of planar drift chambers (CMU), capable of
detecting muons with $p_T > 1.4\,\rm{GeV}/c$~\cite{Ascoli:1987av}.  Four
additional layers of planar drift chambers (CMP) behind another 60~cm of steel detect muons
with $p_T > 2.8$ GeV/$c$~\cite{Dorigo:2000ip}.  These two systems cover the same
central pseudorapidity region with $|\eta| \leq 0.6$.  Muons that
exit the calorimeters at $ 0.6 \leq |\eta| \leq 1.0$ are detected by
the CMX system of four drift layers.  Muon candidates are then
identified as isolated tracks which extrapolate to line segments or
``stubs'' in the muon subdetectors.  A track that is linked to
both CMU and CMP stubs is called a CMUP muon.

The missing transverse energy (\MET)\ is a reconstructed quantity
that is
defined as the absolute value of the opposite of the vector sum
of all calorimeter tower energy depositions projected on the transverse
plane.  It is often used as a measure of the sum
of the transverse momenta of the particles that escape detection, most
notably neutrinos.  To be more readily interpretable as such, the raw
\MET\ vector is adjusted for corrected jet energies and the muon momentum is 
also added for any minimum ionizing high-$p_T$ muon found in the event.

Muon and electron candidates used in this analysis are identified during data taking with the CDF trigger 
system, a three-level filter, with tracking information available at the first level~\cite{Thomson:2002xp}.
Events used in this analysis have all passed the high-energy electron
or muon trigger selection.  The first stage of the central electron
trigger (CEM) requires a track with $p_T > 8$~GeV/$c$ pointing to a tower with
$E_T > 8$~GeV and $E_{\mathrm{HAD}}/E_{\mathrm{EM}}<0.125$. As appropriate for selecting $W$-decay electrons, the plug electron trigger (MET+PEM) requires a tower with $E_T > 8$~GeV, $E_{\mathrm{HAD}}/E_{\mathrm{EM}}<0.125$ and the missing transverse energy \MET $> 15$ GeV.  The first
stage of the muon trigger requires a track with $p_T > 4$~GeV/$c$
(CMUP) or 8~GeV/$c$ (CMX) pointing to a muon stub.  A complete lepton
reconstruction is performed online in the final trigger stage, where
we require $E_T > 18\,\mathrm{GeV}$ for central electrons (CEM), $E_T > 18\,\mathrm{GeV}$ and \MET $> 20\, \mathrm{GeV}$ for plug electron (MET+PEM) and $p_T > 18\,\mathrm{GeV}/c$ for muons (CMUP,CMX).

\section{Event Selection}
\label{sec:eventSelection}

The results presented here use data collected between February 2002
and May 2007.  The data collected using the CEM, CMUP and MET+PEM
triggers correspond to $1.92 \pm 0.12$~fb$^{-1}$ of integrated luminosity, while the data from the
CMX trigger corresponds to $1.88 \pm 0.11$~fb$^{-1}$.  


 
%

The observable final state from the $WH\rightarrow \ell\nu b\bar b$
signal consists of two $b$-jets from Higgs decay, while the decay of $W$ yields the
high-$p_T$ lepton and large missing transverse energy from the neutrino. 
Therefore events are considered as $WH$ candidates only if they have exactly one
lepton candidate~\cite{Acosta:2004uq}, with
$E_T>20\,\mathrm{GeV}$ for electrons or $p_T>20\,\mathrm{GeV}/c$ for
muons. Because the lepton from a leptonic $W$ decay is well-isolated from the 
rest of event, the cone of $\Delta R=0.4$ surrounding the lepton
is required to contain less than 10\% of the lepton energy.  A primary event vertex
position is calculated by fitting a subset of particle tracks that
are consistent with having come from the beamline.  The closest distance $z_0$ along the beam line
between the primary event vertex and the lepton track must be
less than 5~cm to ensure the lepton and the jets come from
the same hard interaction.  Some leptonic $Z$ decays would mimic the
single-lepton signature if one of the leptons is missed. Therefore $Z$ boson candidate events are rejected if a track, an EM cluster or a jet together with the primary lepton forms an invariant mass between 76 and 106 GeV$/c^2$. 
The selected events are
required to have \MET\ greater than 20 GeV, to be consistent with the presence of a neutrino from $W$ decay.

In the plug region, we also require a high-$p_{T}$ isolated lepton
candidate with $E_{T} > 20$ GeV, with the same selection criteria
as for the central region. In addition, because the background contamination from multijet processes is
higher in the forward region, we impose stricter criteria on the missing
transverse energy, which improves the rejection of hadronic background events by 
a factor of four while only reducing the signal efficiency by 20\%.
We require that  $\mathrm{MET_{sig}} > 2$, \MET $>$ 25
GeV and \MET $> 45$~GeV when the \MET\ is pointing 
close to a jet, and large
transverse mass of the reconstructed $W$, $M_{T} (W) > 20$ GeV$/c^2$. Here,
$\mathrm{MET_{sig}}$ is defined as the ratio of \MET\ to the square root of a weighted sum of jet $E_T$ with factors
correlated with mismeasurement, such as angles between the \MET\ and
the jet and size of jet energy corrections~\cite{Peiffer:2008zza},
and $M_{T} (W)$ is defined
as follows:
\begin{equation}
 M_{T} (W) = \sqrt{2(p_{T}^{lep}\ /\!\!\!\!\!E_{T}-{\vec p_{T}^{lep}}\!\cdot/\!\!\!\!\! {\vec E_{T}})}, \label{eq:mtw}
\end{equation}

The $WH$ signal includes two jets originating from $H\rightarrow
b\bar{b}$ decays; these jets are expected to have large transverse
energy.  The jets are required to be in the pseudorapidity range
covered by the silicon detector so that secondary vertices from $b$
decays can be reconstructed.  Specifically, we require the jets to
satisfy $E_T>20$~GeV and $|\eta|<2.0$ and use loose jets ($E_{T} > 12$ GeV and
  $|\eta| < 2.4$) as input for the neural network discriminant discussed in Sec.~\ref{sec:neuralnet}. The search for $WH\rightarrow
\ell\nu b\bar b$ is performed in the sample of events with $W$+
exactly 2 jets; however, samples of events with $W$+1,3,$\geq$4 jets
are used to cross-check the background modeling.
   
To increase the signal purity of the $W$+2 jets events, at least one jet must 
be $b$-tagged by the {\sc secvtx} algorithm as explained in the subsequent sections. Three exclusive $b$-tagged event 
categories 
are considered. The first category (ST+ST) is for events where there are 
two {\sc secvtx} $b$-tagged jets. The second category (ST+JP) consists of 
events where only one of 
the jets is $b$-tagged by the {\sc secvtx} and the second jet is $b$-tagged 
by jet probability.
The third category (ST with NN filter) contains events where only one of the 
jets is $b$-tagged by the {\sc secvtx} and also passes the neural network
 $b$-tagging filter. 

These three categories are selected exclusively and the tightest ST+ST category is preferentially selected. Subsequently the ST+JP category is considered if the ST+ST selection failed and finally ST with NN filter category is considered if both double $b$-tagged selections failed. 
  


\section{$b$ jet identification algorithms}
\label{sec:btag}
Multijet final states have dominant contributions from QCD light-flavor jet production, but the standard model Higgs boson decays
predominantly to bottom quark pairs. Correctly identifying the $b$
quark jets helps to remove most of the QCD background. 

The $b$-quark has a relatively long lifetime, 
and $b$ hadrons formed during the hadronization of the initial $b$ quark can
travel a significant distance before decaying into a collection of
lighter hadrons. Jets containing a $b$-quark decay can be reconstructed by identifying
tracks significantly displaced from the
$p\bar{p}$ interaction point (primary vertex).

In this analysis, we employ three $b$-identification algorithms 
to optimize the selection of $b$-quark jets.

\subsection{Secondary Vertex $b$-Tagging}
\label{sec:secvtx}

The {\sc secvtx} $b$-tagging algorithm is applied to each jet in the
event, using only the tracks which are within $\eta$-$\phi$ distance
of $\Delta R=0.4$ of the jet direction.  Displaced tracks in jets are
used for the {\sc secvtx} reconstruction and are distinguished by a
large impact parameter significance ($|d_0/\sigma_{d_0}|$), where $d_0$
and $\sigma_{d_0}$ are the impact parameter and the total uncertainty
from tracking and beam position measurements, respectively.  Secondary vertices are
reconstructed with a two-pass approach which tests for high-quality vertices 
in the first pass and allows lower-quality vertices in the second pass.
In pass 1, at least three tracks are required to pass loose selection criteria
($p_T>0.5\,\mathrm{GeV}/c$, $|d_0/\sigma_{d_0}|>2.0$), and a secondary vertex is
fit from the selected tracks.  One of the tracks used in the
reconstruction is required to have $p_T>1.0\,\mathrm{GeV}/c$.  If pass 1 fails,
then a vertex is sought in pass 2 from at least two tracks satisfying
tight selection criteria ($p_T>1.0\,\mathrm{GeV}/c$, $|d_0/\sigma_{d_0}|>3.5$
and one of the pass 2 tracks must have $p_T>1.5\,\mathrm{GeV}/c$).  If either
pass is successful, the transverse distance ($L_{xy}$) from the primary
vertex of the event is calculated along with the associated
uncertainty.  This uncertainty $\sigma_{L_{xy}}$ includes
the uncertainty on the primary vertex position.  Finally jets are
tagged positively or negatively depending on the $L_{xy}$ significance
($L_{xy}/\sigma_{L_{xy}}$):
\begin{eqnarray} L_{xy}/\sigma_{ L_{xy}} &\geq& 7.5\ \ \quad
{\mathrm{(positive\ tag)}} \label{eq:postag} \\ L_{xy}/\sigma_{
L_{xy}} &\leq& -7.5 \quad {\mathrm{(negative\ tag)}} \label{eq:negtag}
\end{eqnarray}
This value has been tuned for optimum efficiency and purity in
simulated $b$-jet samples from decays of top quarks.  The energy
spectrum for those jets is similar to the spectrum for $b$ jets from
decays of Higgs bosons.

The sign of $L_{xy}$ indicates the position of the secondary vertex
with respect to the primary vertex along the direction of the jet.  If
the angle between the jet axis and the vector pointing from the primary
vertex to the secondary vertex is less than $\pi/2$, $L_{xy}$ is
positively defined; otherwise, it is negative. If $L_{xy}$ is
positive, the secondary vertex points towards the direction of the
jet, as in true $b$ hadron decays.  For negative $L_{xy}$ the
secondary vertex points away from the jet; this may happen as a result
of mismeasured tracks.
In order to reject secondary vertices due to
material interaction, the algorithm vetoes two-track vertices found
between 1.2 and 1.5 cm from the center of the silicon detector (the
inner radius of the beampipe and the outer radius of the innermost
silicon layer being within this range). All vertices more than 2.5 cm
from the center are rejected because $b$-jets in $WH$ events travel 3.5 mm on average.

The negative tags are useful for evaluating the rate of false positive
tags, which are identified as ``mistags'' in the background estimates.
Mismeasurements are expected to occur randomly; therefore the $L_{xy}$
distribution of fake tags is expected to be symmetric around
zero. Simulated events are used to correct a small asymmetry due to true long-lived particles in light-flavor jets and to interactions in the detector material.

The efficiency for identifying a secondary vertex is
different in simulation and data. We measure an
efficiency correction factor, which is defined as the ratio of the observed to the simulated
efficiencies, to be $0.95\pm0.04$ in a sample of high-$E_T$ jets
enriched in $b$ jets by requiring a soft lepton ($p_T >
8\,\mathrm{GeV}/c$) from semileptonic heavy quark
decays~\cite{Acosta:2004hw}.

\subsection{Neural Network $b$-Tagging Filter}
\label{sec:nnbtag}

The sample tagged by the {\sc secvtx}
algorithm still has significant contamination from falsely-tagged
light-flavor or gluon jets and the misidentification of $c$ quarks as
$b$-jets~\cite{Sal:1}. This search uses a multivariate neural network (NN)
technique to improve the {\sc secvtx} tagging purity~\cite{Aaltonen:2007wx,Aaltonen:2008zx}.


The neural network used in this article employs the {\sc
jetnet}~\cite{Peterson:1993nk} package. The tagger is designed with two
neural networks in series. The $b-l$ network is trained to separate
$b$-jets from light-quark jets ($l$-jets), and the $b-c$ network is
trained to separate $b$-jets from $c$-jets. Jets that pass a cut on
both of the neural network outputs are accepted by the tagger. These
neural networks are trained and applied only to jets that are
already tagged by the {\sc secvtx} algorithm.  The current neural network
$b$-tagging filter is tuned to increase the purity of the {\sc secvtx}
$b$-tagged jets.


The neural networks take as input 16 variables chosen to distinguish $b$-quark jets from $c$- and $l$-quark 
jets by means of their higher track multiplicity, larger invariant mass, longer lifetime, and a harder fragmentation function than
$c$- and $l$-quark jets. The track parameters and $L_{xy}$ significance
are good discriminators for $b$-jets. The sum of transverse momentum $p_T^{\mathrm{vtx}}$ and mass $M_{\mathrm{vtx}}$ of 
the tracks associated with the displaced vertex are useful variables
for identifying $l$-jets; however $c$-jets have $p_T$ spectra
similar to $b$-jets. Pseudo-$c\tau$ ($L_{xy}\times M_{\mathrm{vtx}}/p_T^{\mathrm{vtx}}$), the vertex fit $\chi^2$, and the
track-based probability for a jet to originate away from the primary vertex are
the best discriminators for $b$-jets. 
 
The neural network $b$-tagging filter is validated by comparing the performance on 
a $b$-enriched sample of {\sc secvtx} tagged heavy-flavor jets from events with an electron
candidate with $E_T>8\,\mathrm{GeV}$ and from the
corresponding Monte Carlo sample. A good agreement is found in neural network $b$-tagging filter
performance between data and Monte Carlo~\cite{Aaltonen:2007wx,Aaltonen:2008zx}.
 
The output of the neural network is a value ranging from 0 to 1. We cut on this value so that
we reject 65\% of light-flavor jets and about 50\% of the $c$ jets while
keeping 90\% of $b$-jets which were tagged by {\sc secvtx}.
The data-to-Monte-Carlo correction factor,
measured from the electron sample, is $0.97\pm0.02$. Note that this is
an additional correction factor with respect to the {\sc secvtx} efficiency scale
factor because all of the jets under consideration have already been
tagged by {\sc secvtx}.  

\subsection{Jet Probability $b$-tagging}
\label{sec:jetprob}
The jet probability
$b$-tagging algorithm differs from {\sc secvtx} in that it employs the signed impact
parameters (and their uncertainties) of all tracks within a jet and calculates the
probability that the jet was produced at a position consistent with
the primary vertex.  The signed impact parameter of a track is defined as positive if the angle
between the jet direction and the line joining the primary vertex to the point of closest approach of the track is 
less than 90 degrees, and it is defined as negative otherwise.  
A feature of this algorithm is that the
$b$-tagging is performed using a continuous variable instead of a
discrete object like a reconstructed secondary vertex.

For a light-quark jet, most particles should originate from the primary vertex. Because of the finite tracking resolution, 
charged particle tracks can be reconstructed with a nonzero impact parameter and have an equal probability 
to be either positively or negatively signed.
Since a long-lived particle will travel some distance along the jet direction before decaying, 
its decay products will preferentially have positively signed impact parameters. 



To calculate the track probability, the tracking resolution can be
extracted by fitting the negative side of the signed
impact parameter distribution from the inclusive jet data which is dominated by prompt jets.  
Tracks are sorted into different
categories ($\eta$, $p_{T}$ of tracks, and quality of silicon detector
hits) to parametrize their properties. 
To minimize the contribution
from badly measured tracks with large reconstructed impact parameters,
the distribution of a related quantity, the signed impact parameter
significance (ratio of the signed impact parameter to its
uncertainty) is parametrized for each track category.  Each track is required to satisfy the
quality criteria of $p_{T} > 0.5 \textrm{ GeV}/c$ and a minimum number of hits 
in the tracking detector.
To calculate the jet probability from tracks, at least two tracks with positive impact parameter are
required.  By definition, the jet probability
distribution should be flat between 0 and 1 for jets having only prompt tracks and peaked near zero for heavy flavor jets having large positively signed impact parameter. Tracks
with a negative impact parameter are used to define a negative
jet probability, which is used to check the algorithm and to estimate the
misidentification rate. A jet is considered as tagged if the jet probability is less than 5\%. We choose this value such that we reject 95\% of light-flavor jets while keeping 60\% of $b$-jets. The difference in efficiency between the
simulated and observed data is taken into account as a scale
factor. We measure the scale factor to be 0.85$\pm$0.07 in a sample of
high-$E_{T}$ jets enriched in $b$ jets by requiring a soft lepton from
semileptonic heavy-flavor decay. A more detailed description of the scale
factor estimation is given in Ref.~\cite{Abulencia:2006kv}.

\section{Background}\label{sec:bkg}

The final state signature of $WH\rightarrow \ell\nu b\bar b$ production
can be mimicked by other processes.  The dominant background
processes are $W$+jets production, $t\bar t$ production, and
non-$W$ QCD multijet production.  Several electroweak production
processes also contribute but with smaller rates.
In the following subsections the contribution from each
background source is discussed in detail. These background
estimations are based on the same strategies used in the previous analysis~\cite{Aaltonen:2007wx,Aaltonen:2008zx}.
The quantitative background estimates can be found in Sec.~\ref{sec:results}.

\subsection{Non-$W$ QCD Multijet}
\label{sec:nonW}

Events from QCD multijet production sometimes mimic the
$W$-boson signature due to fake leptons and fake \MET.  A non-$W$
lepton is reconstructed when a jet passes the lepton selection
criteria
or a heavy-flavor jet produces leptons via semileptonic decay.
Non-$W$ \MET\ can result from mismeasurements of energy or
semileptonic decays of heavy-flavor quarks. Since the \MET\ mismeasurement
is usually not well modeled in detector simulation, we directly 
estimate the contribution of non-$W$ events from the lepton+jets data
before $b$-tagging is applied, known as the pretag sample.

Generally, the bulk of non-$W$ events are characterized by a
nonisolated lepton and small \MET.  The lepton isolation $I$ is defined
as the ratio of the sum of the calorimeter energy inside a cone of  $\Delta R = 0.4$ around the lepton direction, 
excluding the lepton energy, and the lepton energy itself.
The quantity $I$ is
small if the lepton is well-isolated from the rest of the event, as
typified by a true leptonic $W$ decay.  This feature is used to
extrapolate the expected non-$W$ contribution into our signal region,
namely, small $I$ and large \MET.  
In extracting the
non-$W$ background contribution from data, we make the following two
assumptions: lepton isolation and \MET\ are uncorrelated in non-$W$
events, and the $b$-tagging rate is not dependent on \MET\ in non-$W$
events. The level at which these assumptions are justified determines
the assigned uncertainty.  
The contributions from $t\bar{t}$ and $W$+jets events are subtracted 
according to the calculated cross sections for those processes.

To validate the method and estimate the relevant systematic
uncertainties, we vary the boundaries of the signal and background
regions.
The observed deviations imply a 25\% systematic
uncertainty in the non-$W$ background yield, assigned conservatively
for both the pretag and tagged estimates~\cite{Aaltonen:2007wx,Aaltonen:2008zx}.


The contribution of non-$W$ background passing the NN $b$-tagging filter is determined by using data in the background region ($I>0.2$ and \MET$>$20~GeV), which has
event kinematics similar to non-$W$ events in the signal region
because lepton isolation is the only difference between the two
regions. The non-$W$ estimate after applying the NN $b$-tagging filter is scaled by the ratio of the events before NN $b$-tagging filter to the events 
after NN $b$-tagging filter in this background region; this assumes the NN
$b$-tagging filter is uncorrelated with the isolation.

The non-$W$ estimate for events with at least two $b$-tags is
obtained by measuring the ratio of the number of events with at least
one $b$-tag to the number with at least two $b$-tags in the background region and
applying the ratio to the estimate of tagged non-$W$ events in the
signal region.

For the plug region, the MET+PEM trigger is used, which means
the above method is not valid, because of the 
\MET\ trigger bias. The \MET\ distribution shape difference between
non-$W$ background and other backgrounds is therefore used to measure the
amount of non-$W$ background. To model the \MET\ distribution shape in the non-$W$ events, the control samples
with electron candidates which failed at least two of our standard lepton 
identification criteria are used. We perform a likelihood fit of the \MET\ distribution for observed data using the non-$W$ and other background shapes~\cite{Peiffer:2008zza}.

\subsection{$W$ + Mistagged Jets}
The rate at which {\sc secvtx} falsely tags light-flavor jets is
derived from inclusive jet samples by counting the number of negative mistags. 
These are defined as jets tagged with an unphysical negative decay length, resulting from imperfect tracking resolution.
They are assumed to be a good estimate of falsely tagged
jets, independent to first order of heavy-flavor content in the
generic jet sample and parametrized as a function of $\eta$, number of vertices, jet $E_T$, track multiplicity, z 
position of primary vertex, and total event $E_T$ scalar sum. 
To estimate the positive mistag rate, the negative mistag rate is corrected for the enhancement of positive tags due to real 
secondary vertices in light-flavor jets and to interactions in the detector material. 
This factor is measured in an inclusive jet sample by fitting the asymmetry 
in the vertex mass distribution of positive tags over negative tags~\cite{Abulencia:2006in}. 
The systematic uncertainty on the mistag rate is estimated from the differences between observed and expected negative tags 
when applying the same parametrization to different jet samples. 
  

The mistag rate per jet is applied to
$W$+jets events. The total estimate is corrected for the
non-$W$ QCD fraction and also for the top-quark contributions to the
pretag sample.  To estimate the mistag contribution in NN-tagged
events, we apply the light flavor rejection factor of the NN filter
$0.35\pm0.05$ as measured using light-flavor jets from various data
and simulated samples. To estimate the mistag contribution in double
tagged events, we apply the mistag rate to all untagged jets in $W$ +
1 $b$-tagged jet events. This method accounts for the case of one real $b$-tag plus one mistag and also for the double-mistag case in 
which both $b$-tagged jets are not due to $b$-hadron decays.


For jet probability $b$-tagging, the mistag rate is
derived from inclusive jet samples as a function of $\eta$, $z$ position
of the primary vertex, jet $E_{T}$, track multiplicity, number of vertices, and total event
scalar $E_{T}$. The mistag rate probabilities are derived from jets with negatively signed impact parameters and applied in 
the same way as for the {\sc secvtx} algorithm~\cite{Abulencia:2006kv}.

\subsection{$W$+Heavy Flavor}

The $Wb\bar{b}$, $Wc\bar{c}$, and $Wc$ processes are major background
sources after the requirement of $b$-tagging. 
Large theoretical uncertainties exist for the overall
normalization because current Monte Carlo
event generators can generate $W$+heavy-flavor events only to leading order.
Consequently, the rates for these processes are normalized to data. 
The contribution from true heavy-flavor production in
$W$+jets events is determined from measurements of the heavy-flavor
event fraction in $W+$jets events and the $b$-tagging efficiency for
those events, as explained below.

The fraction of $W$+jets events produced with heavy-flavor jets has
been studied extensively using an {\sc alpgen}~+~{\sc pythia}
combination of Monte Carlo
simulations~\cite{Mangano:2002ea,Corcella:2001wc}.  Calculations of the
heavy-flavor fraction in {\sc alpgen} have been calibrated using a jet
data sample, and measurements indicate that a scaling factor of $1.4\pm0.4$
is necessary to make the heavy-flavor production in Monte Carlo match
the production in $W$+1 jet events. The final
results obtained for heavy-flavor fractions are shown in
Table~\ref{tbl:HFfrac}.  

For the tagged $W$+heavy flavor (HF) background estimate, the heavy-flavor fractions
and tagging rates given in Tables~\ref{tbl:HFfrac}
and~\ref{tbl:HF_tageff} are multiplied by the number of pretag
$W$+jets candidate events ($N_{\rm pretag}$) in data, after correction for the
contribution of non-$W$ ($f_{{\rm non}-W}$), $t\bar{t}$ and other background events to the pretag sample.
 The $W$+heavy flavor background contribution is obtained by the following
relation:
\begin{widetext}
\begin{equation}
  N_{W+{\rm HF}} = f_{{\rm HF}} \cdot \epsilon _{{\rm tag}} \cdot
  \left [ N_{\rm pretag}\cdot (1-f_{{\rm non}-W}) - N_{{\rm TOP}} - N_{{\rm
  EWK}}\right ],
\end{equation}
\end{widetext}
where $f_{HF}$ is the heavy-flavor fraction, $\epsilon_{\rm tag}$ is
the tagging efficiency, $N_{\rm TOP}$ is the expected number of $t\bar
t$ and single top events, and $N_{\rm EWK}$ is the expected background contribution from
$WW$, $WZ$, $ZZ$ and $Z$ boson events.

\begin{table*}
  \begin{center}
  \caption{The heavy-flavor fractions $f_{HF}$, given in percent, for the $W$ +jets events where $1B$, $2B$
   refer to number of taggable $b$-jets in the events, with $1C$, $2C$ for charm jets.
   The results from {\sc alpgen} Monte Carlo have been scaled by the data-derived
   calibration factor of $1.4\pm 0.4$. }
  \begin{tabular}{ccccc}
    \hline \hline
    Jet Multiplicity  & 1 jet & 2 jets & 3 jets & $\geq4$ jets \\
    \hline
 $Wb\bar b$ (1B) (\%)& 1.0 $\pm$  0.4 &  2.0 $\pm$  0.8 &  3.4 $\pm$  1.4 &  4.6 $\pm$  2.0 \\
 $Wb\bar b$ (2B) (\%)&              - &  1.3 $\pm$  0.6 &  2.5 $\pm$  1.0 &  3.1 $\pm$  1.8 \\
 $Wc\bar c$ (1C) (\%)& 7.7 $\pm$  2.4 &  12.2 $\pm$  4.5 &  16.4 $\pm$  5.3 &  18.6 $\pm$  6.9 \\
 $Wc\bar c$ (2C) (\%)&              - &  2.0 $\pm$  0.8 &  4.6 $\pm$  1.8 &  8.4 $\pm$ 3.4 \\
    \hline \hline
  \end{tabular}
   \label{tbl:HFfrac}
  \end{center}
\end{table*}

\begin{table*}
  \begin{center}
\caption{The $b$-tagging efficiencies $\epsilon_{tag}$ in percent for the various $b$-tagging strategies
   on individual $W$+heavy-flavor processes.  Categories $1B$, $2B$
   refer to number of taggable $b$-jets in the events, with similar
   categories for charm jets.  Those numbers include the effect of the
   data-to-Monte Carlo scale factors.}
\begin{tabular}{ccccc}
\hline\hline
   Jet Multiplicity &   1 jet &   2 jets &   3 jets &  $\geq4$ jets \\
 \hline
 \multicolumn{5}{c}{$1$ {\sc secvtx} and NN $b$-tag (\%)}\\
 
 $Wb\bar b$ (1B) & 27.5 $\pm$ 1.5 & 28.1 $\pm$ 1.3 & 26.7 $\pm$ 1.4 & 26.9 $\pm$ 3.7 \\
 $Wb\bar b$ (2B) &            -   & 26.2 $\pm$ 1.2 & 24.1 $\pm$ 1.2 & 22.6 $\pm$ 1.3 \\
 $Wc\bar c$ (1C) &  4.2 $\pm$ 0.2 & 4.6 $\pm$ 0.3 & 4.9 $\pm$ 0.3 & 5.2 $\pm$ 0.7 \\
 $Wc\bar c$ (2C) &            -   & 6.3 $\pm$ 0.4 & 6.6 $\pm$ 0.4 & 6.9 $\pm$ 0.6 \\
\hline
 \multicolumn{5}{c}{$\geq2$ {\sc secvtx} $b$-tag (\%)}\\
 $Wb\bar b$ (2B) &            -   & 16 $\pm$ 2 & 19 $\pm$ 2 & 19 $\pm$ 3 \\
 $Wc\bar c$ (2C) &            -   & 1.0 $\pm$ 0.5 & 2.0 $\pm$ 1.0 & 2.0 $\pm$ 1.0 \\
\hline
 \multicolumn{5}{c}{$1$ {\sc secvtx} + Jet Probability $b$-tag (\%)}\\
 $Wb\bar b$ (2B) &            -   & 10.1 $\pm$ 1.2 & 11.1 $\pm$ 1.3 & 12.4 $\pm$ 1.5 \\
 $Wc\bar c$ (2C) &            -   & 1.6 $\pm$ 0.2 & 2.3 $\pm$ 0.3 & 3.2 $\pm$ 0.4 \\

 \hline\hline
\end{tabular}
\label{tbl:HF_tageff}
  \end{center}
\end{table*}

\subsection{Top and Electroweak Backgrounds}
Production of both single top quarks and top-quark pairs contributes to
the tagged $W$+jets sample. Several electroweak boson production
processes also contribute. $WW$ pairs can decay to a lepton, neutrino
as missing energy, and two jets, one of which may be charm. $WZ$
events can decay to the signal $\ell \nu b\bar b$ or $\ell \nu c\bar c$ final
state. Finally, $Z\rightarrow\tau^+\tau^-$ events can have one
leptonic $\tau$ decay and one hadronic decay. The leptonic $\tau$
decay gives rise to a lepton plus missing transverse energy, while the
hadronic decay yields a narrow jet of hadrons with a nonzero
lifetime.

The normalization of the diboson and top production backgrounds are based
on the theoretical cross sections listed in Table~\ref{tbl:xsec}, the
luminosity, and the acceptance and $b$-tagging efficiency derived from
Monte Carlo
events~\cite{Campbell:2002tg,Acosta:2004uq,Cacciari:2003fi,Harris:2002md}. The
acceptance is corrected for lepton identification, trigger
efficiencies, and the $z$ vertex cut. The tagging efficiency is corrected by the $b$-tagging scale factor.

\begin{table}
  \begin{center}
    \caption{Theoretical
    cross sections and uncertainties for the electroweak and single top
    backgrounds, along with the theoretical cross section for
    $t\bar{t}$ at $m_t = 175\,\mathrm{GeV}/c^2$~\cite{Eidelman:2004wy}. The cross section of $Z^{0}
    \rightarrow \tau^+\tau^-$ is obtained in the dilepton mass range
    $m_{\tau\tau}>30\,\mathrm{GeV}/c^2$ together with a $k$-factor (NLO/LO) of
    1.4~\cite{Abulencia:2007iy} }
    \begin{tabular}{cc}
      \hline \hline
      Background & Theoretical Cross Sections\\ \hline  
	$WW$ & 12.40 $\pm$ 0.25 pb \\ 
	$WZ$ & 3.96 $\pm$ 0.06 pb \\ 
	$ZZ$ & 1.58 $\pm$ 0.05 pb \\ 
	Single top $s$-channel & 0.88 $\pm$ 0.11 pb\\ 
	Single top $t$-channel & 1.98 $\pm$ 0.25 pb\\
    	$Z\rightarrow \tau^+\tau^- $ & 265 $\pm$ 30.0 pb\\ 
	$t\bar{t}$ & 6.7$^{+0.7} _{-0.9}$ pb \\
    \hline\hline
    \end{tabular}
    \label{tbl:xsec}
  \end{center}
\end{table}

\section{Higgs Boson Signal Acceptance}
\label{sec:Acceptance}
\begin{figure}[htbp]
\begin{center}
 \includegraphics[width=0.48\textwidth]{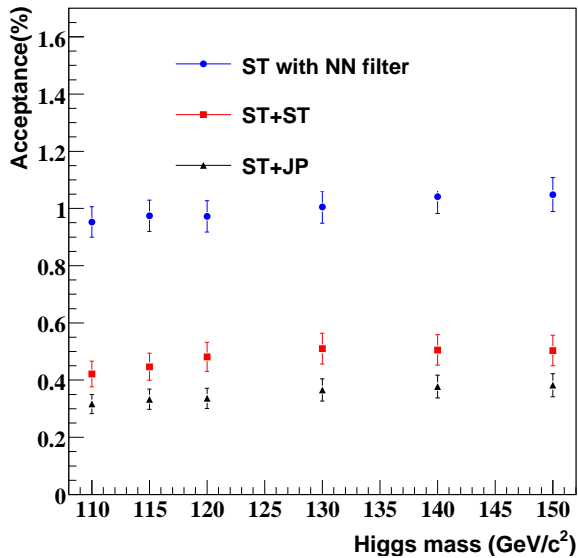}
\end{center}
\caption{The total acceptance $\epsilon$ for the process $WH \rightarrow \ell
 \nu b\bar{b}$ in $W$+2jet bin for the selected $b$-tagging strategies
 as a function of Higgs boson mass.
\label{fig:accwoPHX}}
\end{figure}

The {\sc pythia} Monte Carlo generator is used to generate the signal
samples and study the kinematic properties~\cite{Sjostrand:2000wi}.
Only Higgs boson masses between 110 and $150\,\mathrm{GeV}/c^2$ are
considered because this is the mass region, especially up to $135\,\mathrm{GeV}/c^2$, for which the decay
$H\rightarrow b\bar b$ dominates.  The number of expected
$WH\rightarrow \ell\nu b\bar b$ events $N$ is given by
\begin{equation}
N = \epsilon \cdot \int {\cal{L}} dt \cdot \sigma (p \bar{p}
 \rightarrow WH)\cdot {\cal B}(H \rightarrow b \bar{b}), \label{eq:ExpEvt}
\end{equation}
where $\epsilon$, $\int {\cal{L}}
dt$, $\sigma(p \bar{p} \rightarrow WH)$, and ${\cal B}(H\rightarrow
b\bar{b})$ are the event detection acceptance, integrated luminosity,
production cross section, and branching fraction, respectively.  The
production cross section and branching fraction are calculated to NLO
precision~\cite{Djouadi:1997yw}.  The acceptance
$\epsilon$ is broken down into the
following factors:
\begin{widetext}
\begin{equation}
\epsilon = 
 \epsilon_{z_0} \cdot \epsilon _{b\mathrm{tag}} \cdot \sum_{\ell=e,\mu,\tau } \left( \epsilon_{\mathrm{trigger}} \cdot \epsilon_{\mathrm{lepton\ ID}}
\cdot \epsilon_{\mathrm{kinematics}} \cdot  {\cal B}(W \rightarrow \ell \nu) \right), \label{eq:sigEff}
\end{equation}
\end{widetext}
where $\epsilon_{z_0}$, $\epsilon_{\mathrm{trigger}}$,
$\epsilon_\mathrm{lepton\ ID}$, $\epsilon _{b\mathrm{tag}}$, and
$\epsilon _{\mathrm{kinematics}}$ are efficiencies defined in sequence to meet the
requirements of primary vertex, trigger, lepton identification,
$b$-tagging, and event selection criteria.  The major sources of inefficiency are
the lepton identification, jet kinematics, and $b$-tagging factors;
each has an efficiency between 30 and 45\%.  
The factor $\epsilon_{z_0}$ is obtained using the vertex distribution from the minimum bias data, $\epsilon_{trigger}$ is 
measured using a clean $W\rightarrow l\nu$ data sample, obtained from different triggers after applying more 
stringent offline cuts,
and $\epsilon_\mathrm{lepton\ ID}$ is 
calculated using $Z\rightarrow ll$ observed data and 
Monte Carlo samples. $\epsilon _{b\mathrm{tag}}$ is measured in a $b$-enriched sample from semileptonic heavy flavor decay.
The total signal acceptances are shown in Fig.~\ref{fig:accwoPHX} for the selected $b$-tagging options
as a function of Higgs boson mass. Inclusion of forward electrons in this analysis increased the overall acceptance by 10\%.

The total systematic uncertainty on the acceptance stems from the jet
energy scale, initial and final state radiation (ISR and FSR), lepton
identification, trigger efficiencies, and $b$-tagging scale factor.
A 2\% uncertainty on the lepton identification efficiency is assigned
for each lepton type (CEM electron, PEM electron, CMUP and CMX muon), based on
studies of $Z$ boson events.  For each of the high $p_T$ lepton
triggers, a 1\% uncertainty is measured from backup trigger paths or
$Z$ boson events.  

The initial and final state radiation systematic
uncertainties are estimated by changing the parameters in the Monte Carlo related to ISR
and FSR from nominal values to half or double the
nominal~\cite{Abulencia:2005aj}.  Half of the difference from the two samples is taken as the systematic uncertainty.  The uncertainty in
the incoming parton energies relies on the eigenvectors provided in
the PDF fits.  A NLO version of the PDFs, CTEQ6M, provides a 90\%
confidence interval of each eigenvector~\cite{Pumplin:2002vw}. The
nominal PDF value is reweighted to the 90\% confidence level value,
and the corresponding reweighted acceptance is computed.  The
differences between nominal and reweighted acceptances are added in
quadrature, and the total is assigned as the systematic
uncertainty~\cite{Acosta:2004hw}.

\begin{table*}
\begin{center}
\caption{Systematic uncertainties on signal acceptance for the selected $b$-tagging requirements.}
{\small  
\begin{tabular}{cccccccc}\hline \hline 
                    $b$-tagging category &   LeptonID &   Trigger &   ISR/FSR &    JES &    PDF & $b$-tagging & Total\\ \hline 
  ST with NN filter &  $\sim$ 2\%&    $<$ 1\%&      2.9\%&   2.3\%&   1.2\%&        3.5\%& 5.6\% \\ 
  ST+ST &  $\sim$ 2\%&    $<$ 1\%&      5.2\%&   2.5\%&   2.1\%&        8.4\%& 10.6\% \\ 
  ST+JP &  $\sim$ 2\%&    $<$ 1\%&      4.0\%&   2.8\%&   1.5\%&        9.1\%& 10.5\% \\  
\hline \hline
\end{tabular} 
} 
\label{tbl:Sys2jetx}
\end{center}
\end{table*}

The effect of the jet energy scale uncertainty (JES)~\cite{Bhatti:2005ai} is calculated by shifting the jet energy scale by $\pm 1\sigma$ in the $WH$
Monte Carlo samples.  The deviation from the nominal
acceptance is taken as the systematic uncertainty.  The systematic
uncertainty on the $b$-tagging efficiency is based on the scale
factor uncertainty discussed in Sec.~\ref{sec:secvtx} and~\ref{sec:jetprob}. When the NN $b$-tagging filter is applied, the scale factor uncertainty
is added to that of {\sc secvtx} in quadrature.  The total systematic
uncertainties for the selected $b$-tagging options are summarized in
Table~\ref{tbl:Sys2jetx}.
    
\begin{table*}
\begin{center}
\caption{Expected number of $WH\rightarrow \ell \nu b\bar{b}$ signal
  events in central region for the selected $b$-tagging options. All systematic uncertainties are included.
}
{\small  
\begin{tabular}{ccccccc}\hline \hline 
$b$-tagging category & 110 GeV & 115 GeV & 120 GeV & 130 GeV & 140 GeV & 150 GeV\\ \hline 
Pretag & 9.41$\pm$0.61 & 7.92$\pm$0.52 & 6.35$\pm$0.41 & 3.99$\pm$0.26 & 2.02$\pm$0.13 & 0.78$\pm$0.05\\
ST with NN filter & 2.63$\pm$0.22& 2.20$\pm$0.18& 1.70$\pm$0.14& 1.08$\pm$0.09& 0.55$\pm$0.05&  0.21$\pm$0.02\\ 
ST+ST & 1.18$\pm$0.14& 1.00$\pm$0.12& 0.85$\pm$0.10& 0.55$\pm$0.07& 0.27$\pm$0.03& 0.10$\pm$0.01\\ 
ST+JP & 0.89$\pm$0.11& 0.76$\pm$0.09& 0.60$\pm$0.07& 0.40$\pm$0.05& 0.20$\pm$0.02&  0.08$\pm$0.01\\  
\hline \hline
\end{tabular} 
} 
\label{tbl:ExEvt2woPHX}
\end{center}
\end{table*}

The expected number of 
signal events
is estimated by Eq.~\ref{eq:ExpEvt} at each Higgs boson mass point.  The
expectations for the selected $b$-tagging strategies are shown in
Table~\ref{tbl:ExEvt2woPHX}.


\section{Neural Network Discriminant}
\label{sec:neuralnet}
To further improve the signal to background discrimination after event
selection, we employ a neural network~\cite{Peterson:1993nk} trained on a
variety of kinematic variables to distinguish the $W$+Higgs events from the backgrounds.

The neural network on the samples of simulated events using a mixture of 
50\% signal and 50\% of backgrounds is trained. The background composition is chosen 
to have equal amounts of $Wb\bar{b}$, $t\bar t$, and single top, which provides a maximum sensitivity over
a wide range of input conditions. 

To optimize the neural network structure, we use an iterative procedure to determine the
configuration that best discriminates signal from the background, and
uses a minimal number of input discriminants.  This is done by
first determining the best one-variable neural network from a list of 76 possible
variables, based on the kinematic distributions of the two jets, lepton, and
\MET\ in the events (including correlations among these objects).  
The optimization algorithm keeps this variable
as an input and then loops over all other variables to determine the
best two-variable neural network. The best N-variable neural network is finally selected
once the (N+1)-variable neural network shows less than 0.5 \% improvement.
The criteria for comparing neural networks is the testing error defined by
how often a neural network with a given configuration incorrectly classifies
signal and background events.
This optimization of the neural network structure was done for Higgs mass of 120 GeV/c$^{2}$, and we use the same structure to train separate neural networks 
for Higgs masses of 110, 115, 120, 130, 140, and 150 GeV/c$^{2}$.
Retraining neural networks with different signal masses
keeps the neural network improvement almost constant as a function of the Higgs mass. 

Our neural network configuration has 6 input variables, 11 hidden nodes,
and 1 output node.  The output of the neural network has a value from 0 to 1 that will provide discrimination between the signal and background hypotheses. The signal-like events yield a high neural network value and the background like events yield a low neural network value.

The 6 optimal inputs are as follows:

\begin{itemize}
  \item{$M_{jj+}$: the invariant mass calculated from
  the two jets in the $W$+2jets event.  Furthermore, if
  there are additional loose jets present, the loose jet that is closest to one of the two
  jets is included in this invariant mass calculation, if the
  separation between that loose jet and one of the jets is $\Delta R < 0.9$. This definition gives greater discriminating power for instances where the selected $b$-jet spreads outside the cone.}
  \item{$\sum{E_T}$(Loose Jets): the scalar sum of
   the transverse energies of the loose jets.}
  \item{$p_T$ Imbalance: the difference
  between the scalar sum of the transverse momenta of all measured
  objects and the \MET.  Specifically, it is calculated as $
  P_T(jet_{1}) + P_T(jet_{2}) + P_T(lep) -$ \MET.}
  \item{$M_{l\nu j}^{min}$: the invariant mass of the lepton,
  \MET, and one of the two jets, where the jet is chosen to give the
  minimum invariant mass.  For this quantity, the $p_z$ component of the
  neutrino is ignored.}
  \item{$\Delta R$(lepton-$\nu_{max}$): the $\Delta R$
  separation between the lepton and the neutrino, where the $p_z$ of
  the neutrino is taken by choosing the largest $|p_z|$ of solutions from the
  quadratic equations for the $W$ mass (80.42 GeV/c$^{2}$) constraint.}
  \item{$P_T(W+H)$: the norm of the vector sum of transverse momentum of the lepton, \MET\ and two jets.}
\end{itemize}

Distributions of all these quantities are checked for both the pretag and tagged samples. The simulated 
background events match to the real data well. 





\section{Results}\label{sec:results}

\subsection{Counting Results}
\label{sec:bkgsummary}

\begin{figure}[htbp]
  \begin{center}
     \includegraphics*[width=0.48\textwidth]{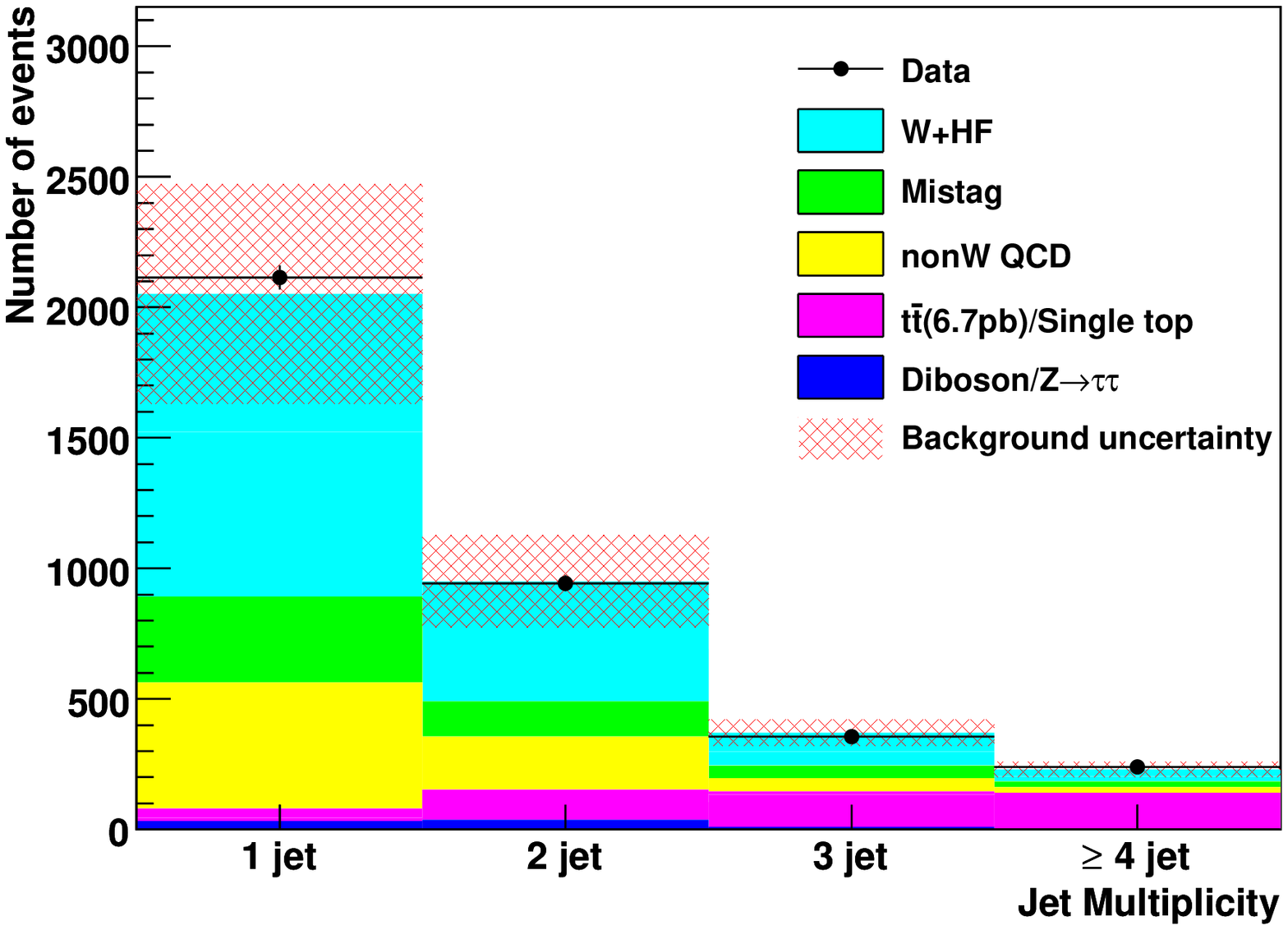}
     \includegraphics*[width=0.48\textwidth]{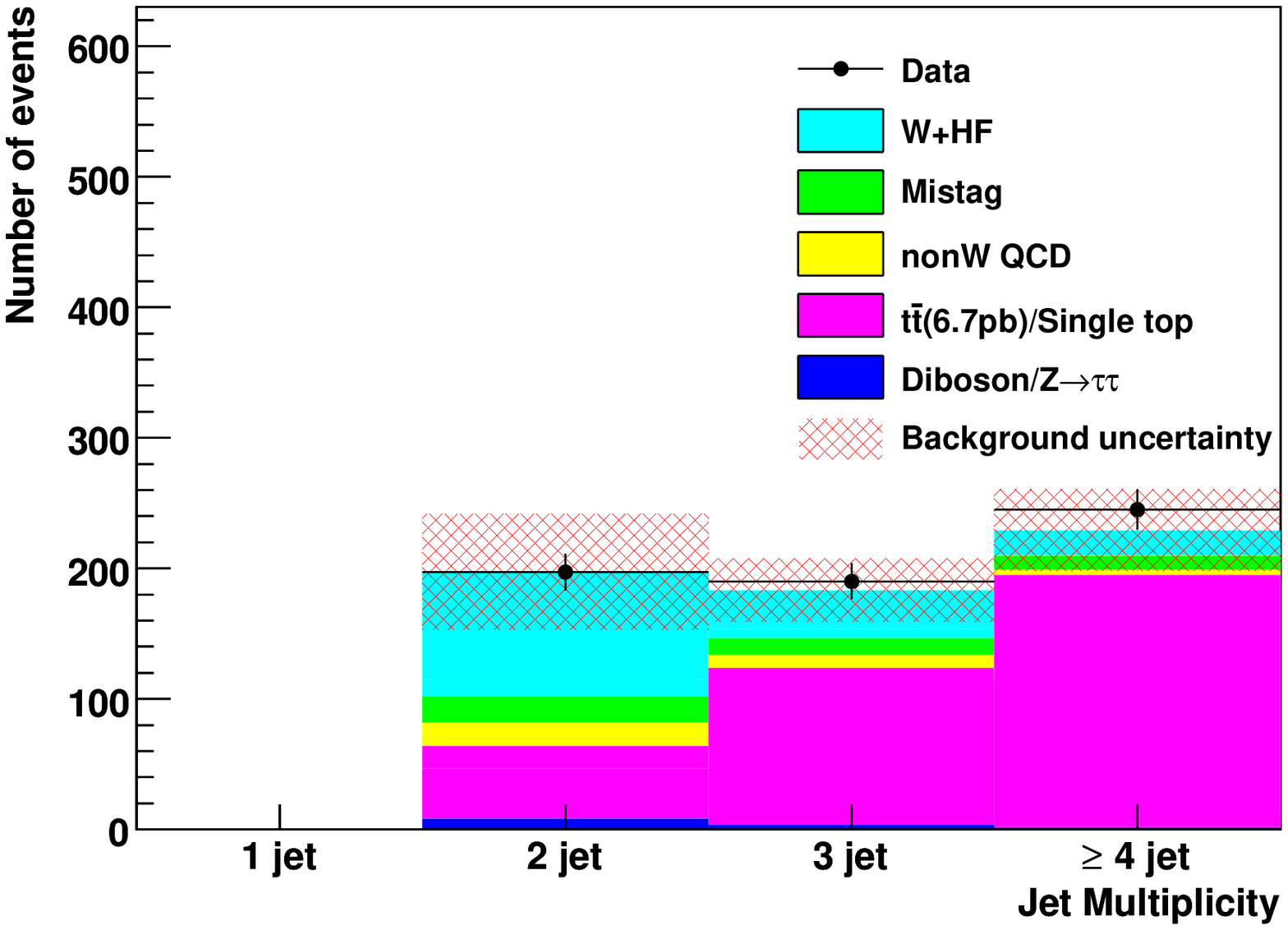}
    \caption{Number of events in the central and plug regions as a function of jet multiplicity
       for events with exactly one {\sc secvtx} $b$-tag applying the NN $b$-tagging filter requirement (left) and for events with 
at least two {\sc secvtx} $b$-tagged jets or one {\sc secvtx} $b$-tagged jet plus one jet probability $b$-tagged jet (right).}
    \label{fig:Njets}
  \end{center}
\end{figure}

The observed number of events in data is compared to the expected background in  Fig.~\ref{fig:Njets}, 
as a function of jet multiplicity. Results are shown for the single and double $b$-tagged categories separately.
Table~\ref{tbl:Njets_all}
shows the composition of $W+2$ jet data in each $b$-tagging category and in  the central lepton and 
plug lepton regions, respectively.

The observed number of events in the data and the standard model background expectations are consistent in each $b$-tagging category.

  \begin{table*}
    \begin{center}
 \caption{Predicted sample composition and observed number of $W+2$ jet events in the central and plug regions with the selected 
$b$-tagging options. }
\begin{tabular}{ccccccc}\hline \hline
                    &  \multicolumn{3}{c}{Central region}   &       \multicolumn{3}{c}{ Plug region} \\ \hline     
       Pretag Events&    \multicolumn{3}{c}{32242} &    \multicolumn{3}{c}{5879} \\ \hline		  
       $b$-tagging  &  ST+ST & ST+JP & ST+NN & ST+ST & ST+JP & ST+NN \\ \hline 
              Mistag&   3.88$\pm$0.35&   11.73$\pm$0.92&  107.1$\pm$9.38&   1.00$\pm$0.18 &  3.18$\pm$0.49& 28.47$\pm$3.30 \\ 
         $Wb\bar{b}$&   37.93$\pm$16.92& 31.15$\pm$14.03& 215.6$\pm$92.34&  7.40$\pm$3.96 &  6.23$\pm$3.37& 43.09$\pm$12.33\\		   
         $Wc\bar{c}$&   2.88$\pm$1.25&   7.87$\pm$3.43&   167.0$\pm$62.14&  0.96$\pm$0.49 & 1.53$\pm$0.81&  33.37$\pm$9.55\\		   
   $t\bar{t}$(6.7pb)&   19.05$\pm$2.92&  15.56$\pm$2.39&  60.68$\pm$9.30&   2.14$\pm$0.34 &  1.79$\pm$0.31&  7.17$\pm$1.00\\		   
    Single top(s-ch)&   6.90$\pm$1.00&   5.14$\pm$0.75&   14.38$\pm$2.09&   0.69$\pm$0.10 &  0.51$\pm$0.08& 1.53$\pm$0.20\\		   
    Single top(t-ch)&   1.60$\pm$0.23&   1.87$\pm$0.27&   29.57$\pm$4.33&   0.22$\pm$0.04 &  0.24$\pm$0.04& 3.54$\pm$0.47\\		   
                $WW$&   0.17$\pm$0.02&   0.93$\pm$0.11&   15.45$\pm$1.91&   0.01$\pm$0.01 &   0.12$\pm$0.04& 3.00$\pm$0.20\\		   
                $WZ$&   2.41$\pm$0.26&   1.84$\pm$0.20&   7.59$\pm$0.81&    0.58$\pm$0.06 &  0.42$\pm$0.05& 1.62$\pm$0.09\\		   
                $ZZ$&   0.06$\pm$0.01&   0.08$\pm$0.01&   0.31$\pm$0.03&    0.00$\pm$0.01 &  0.01$\pm$0.01&0.02$\pm$0.00\\		   
  $Z\rightarrow\tau\tau$& 0.25$\pm$0.04& 1.29$\pm$0.20&   7.27$\pm$1.12&    0.00$\pm$0.01 &   0.01$\pm$0.01&  0.24$\pm$0.03\\		   
         non-$W$ QCD&   5.50$\pm$1.00&   9.55$\pm$1.73&  184.7$\pm$33.04&   1.16$\pm$0.44 &  1.51$\pm$0.55& 18.34$\pm$5.54\\ \hline		   
    Total Background&   80.6$\pm$18.8& 87.0$\pm$18.0& 809.6$\pm$159.4& 14.2$\pm$4.0 & 15.5$\pm$3.6& 140.4$\pm$16.9\\ \hline	   
$WH$ signal (120 GeV)&  0.85$\pm$0.10&   0.60$\pm$0.07&   1.70$\pm$0.14  &   0.09$\pm$0.01 &   0.06$\pm$0.01&  0.20$\pm$0.01\\  \hline	   
     Observed Events&   83&               90&               805&              11&   13& 138 \\ \hline \hline
\end{tabular}
      \label{tbl:Njets_all}
    \end{center}
  \end{table*}

\subsection{Limit on Higgs Boson Production Rate}
\label{sec:limit}

\begin{figure}[htbp]
  \begin{center}
   \includegraphics[width=0.48\textwidth]{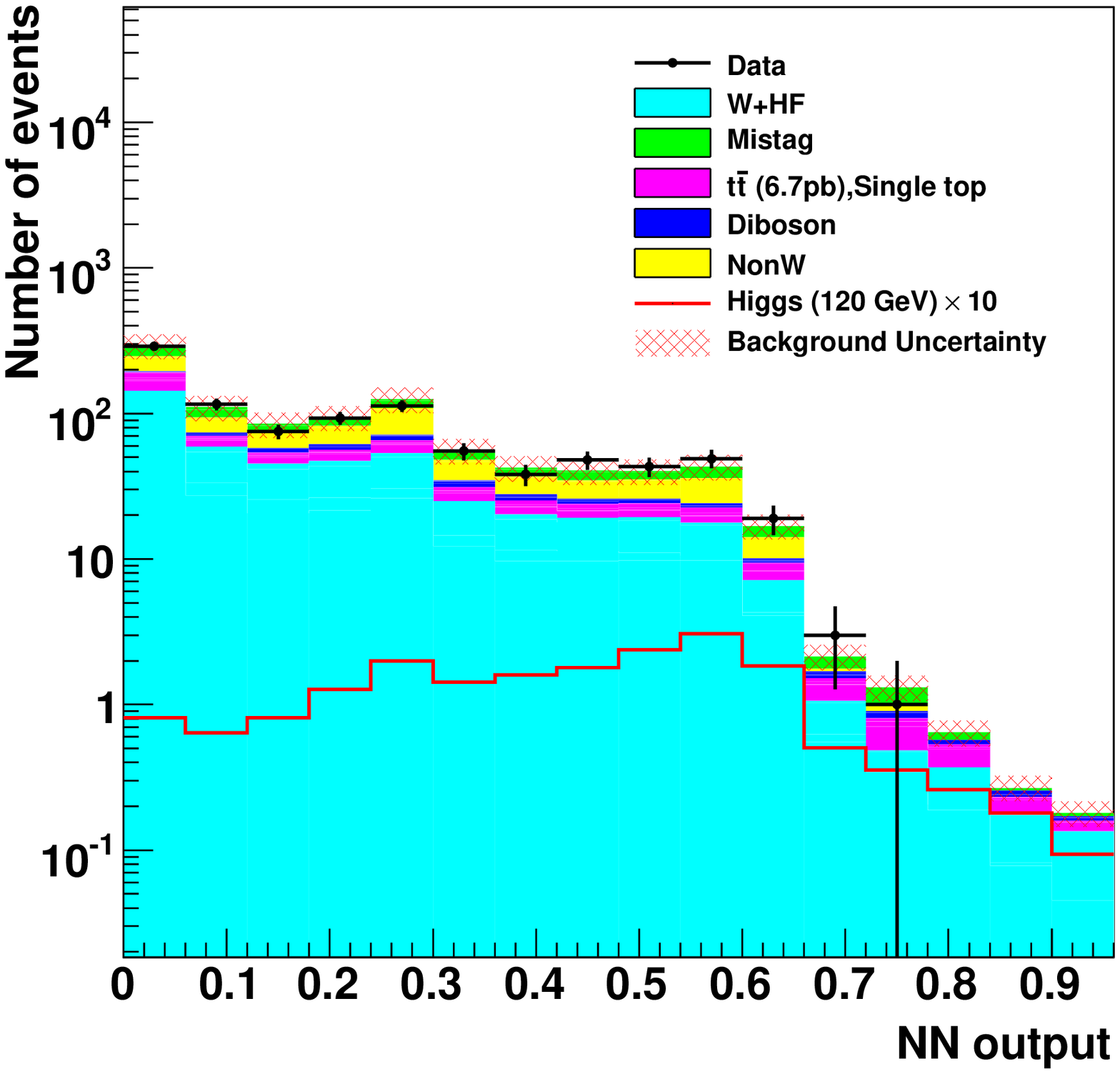}
   \includegraphics[width=0.48\textwidth]{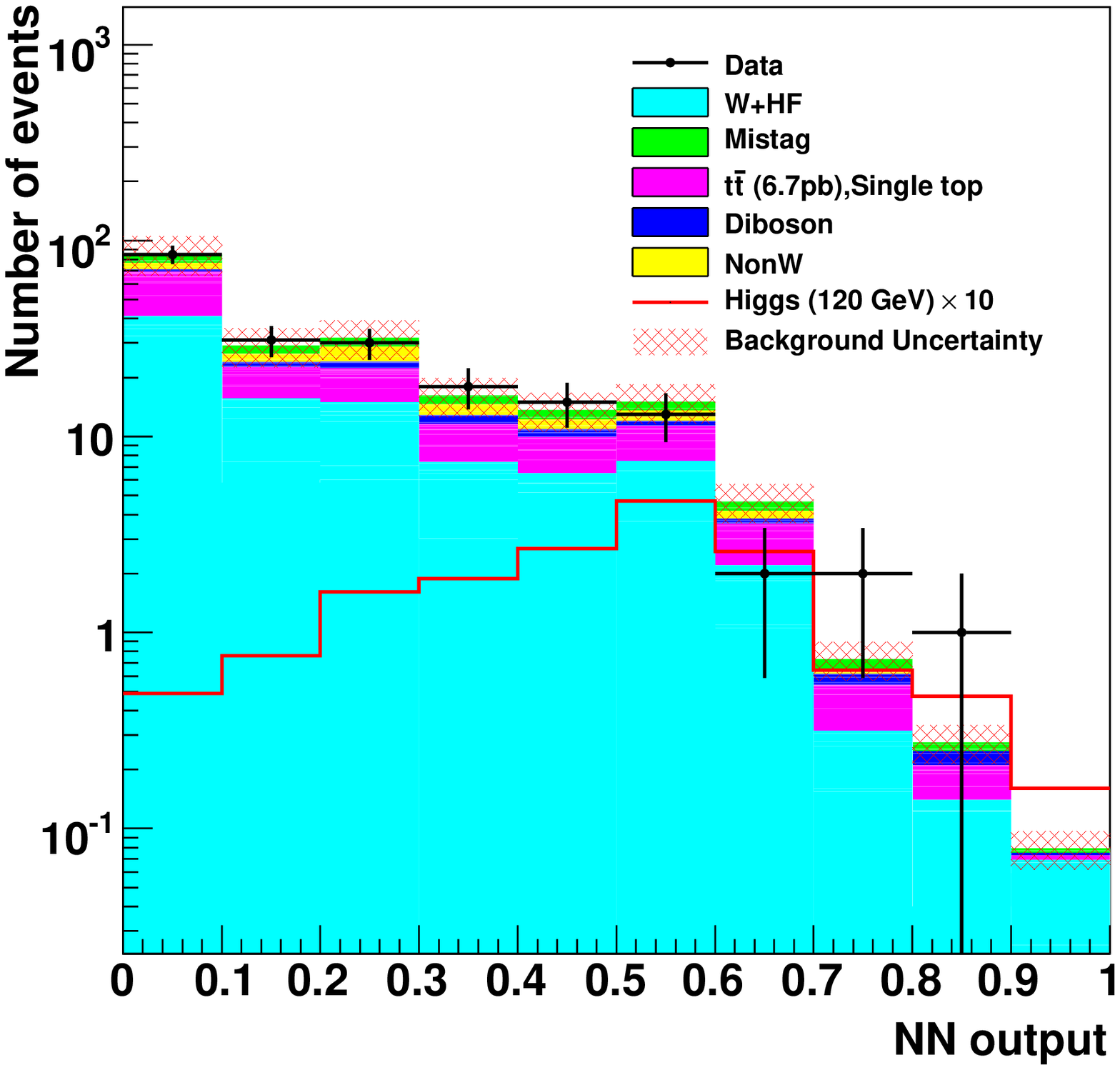}
   \caption{Neural Network output distribution in $W$+2 jets events for exactly one {\sc secvtx} $b$-tagged jet that passes the NN
     $b$-tagging filter (left) and events for ST+ST and ST+JP double $b$-tagging categories (right). 
     The contributions of the various background sources from the central plus plug region
     are shown in histograms while the hatched box represents the background uncertainty. The expected signal 
for a 120 GeV Higgs boson (multiplied by a factor of 10) is shown by the solid line.} 
     \label{fig:nnout}
  \end{center}
\end{figure}

We apply the neural network discriminant discussed in the previous section to the samples of simulated events and obtain the distributions of the network output for all the processes considered. 
The expected distributions are compared to the data observed in 
single and double $b$-tagged categories as shown in Fig.~\ref{fig:nnout}.  
We use a binned likelihood technique to fit the observed neural network output distributions in the three $b$-tagging categories to 
test for the presence of a $WH$ signal.


The number of events in each bin follows the Poisson distribution
\begin{equation}
  P_i (n_i, \mu_i) = \frac{\mu_i ^{n_i} e^{-\mu_i}}{n_i !}
  \label{eq:Poisson} \quad (i=1,2,\cdots, N_{{\rm bin}}),
\end{equation}
where $n_i,\mu_i$, and $N_{\mathrm{bin}}$ represent the number of
observed events in the $i$-th bin, the expectation in the $i$-th bin, and the
total number of bins.  The Higgs production hypothesis is constructed
by setting $\mu_i$ to $\mu_i =	s_i + b_i$,
where $s_i$ and $b_i$ are the number of signal and expected background
events in the $i$-th bin.
This quantity $s_i$ can also be written as a product
\begin{equation}
s_i =  \sigma(p\bar{p}\rightarrow W^{\pm}H)\cdot{\cal B}(H\rightarrow b\bar{b}) \cdot  \epsilon \cdot \int {\mathcal{L}}dt \cdot f_i^{WH},
\end{equation}
where $f_i^{WH}$ is the fraction of the
total signal which lies in the $i$-th bin.  
In this case,
$\sigma(p\bar{p}\rightarrow W^{\pm}H)\cdot{\cal B}(H\rightarrow
b\bar{b})$ is the variable to be extracted from data. 
The likelihoods from the three $b$-tagging categories are multiplied together. 
The systematic uncertainties associated with the pretag acceptance, luminosity, and the $b$-tagging efficiency
scale factor are considered to be fully correlated among the three selection categories. 
Background uncertainties on the heavy-flavor fractions are completely correlated among $W$+HF backgrounds and background uncertainties on $b$-tagging scale factor are also completely correlated among all backgrounds. 
The systematic uncertainties associated with the shape of network output are also studied and found to have a negligible impact on the final results. 
 
Since we observe no excess over the background prediction, 
an upper limit on the Higgs boson production cross section times branching fraction
$\sigma(p\bar{p}\rightarrow W^{\pm}H)\cdot {\cal B}(H\rightarrow
b\bar{b})$ is extracted by using a Bayesian procedure.
We assume a uniform prior probability for $\sigma\cdot {\cal B}$
and integrate the likelihood over all parameters except for
$\sigma\cdot {\cal B}$. The 95\% confidence level upper limits on
$\sigma\cdot{\cal B}$ are obtained by calculating the $95^{th}$
percentile of the resulting distributions.


To evaluate the sensitivity of the analysis, background-only
pseudoexperiments are used to calculate the expected limit in the
absence of Higgs boson production.  Pseudo data are generated by
fluctuating the individual background estimates within their total
uncertainties. The expected limit is defined as the median of the 95\% confidence level upper limits of 
1000 pseudoexperiments. 

\begin{figure}[htbp]
  \begin{center}
     \includegraphics[width=0.48\textwidth]{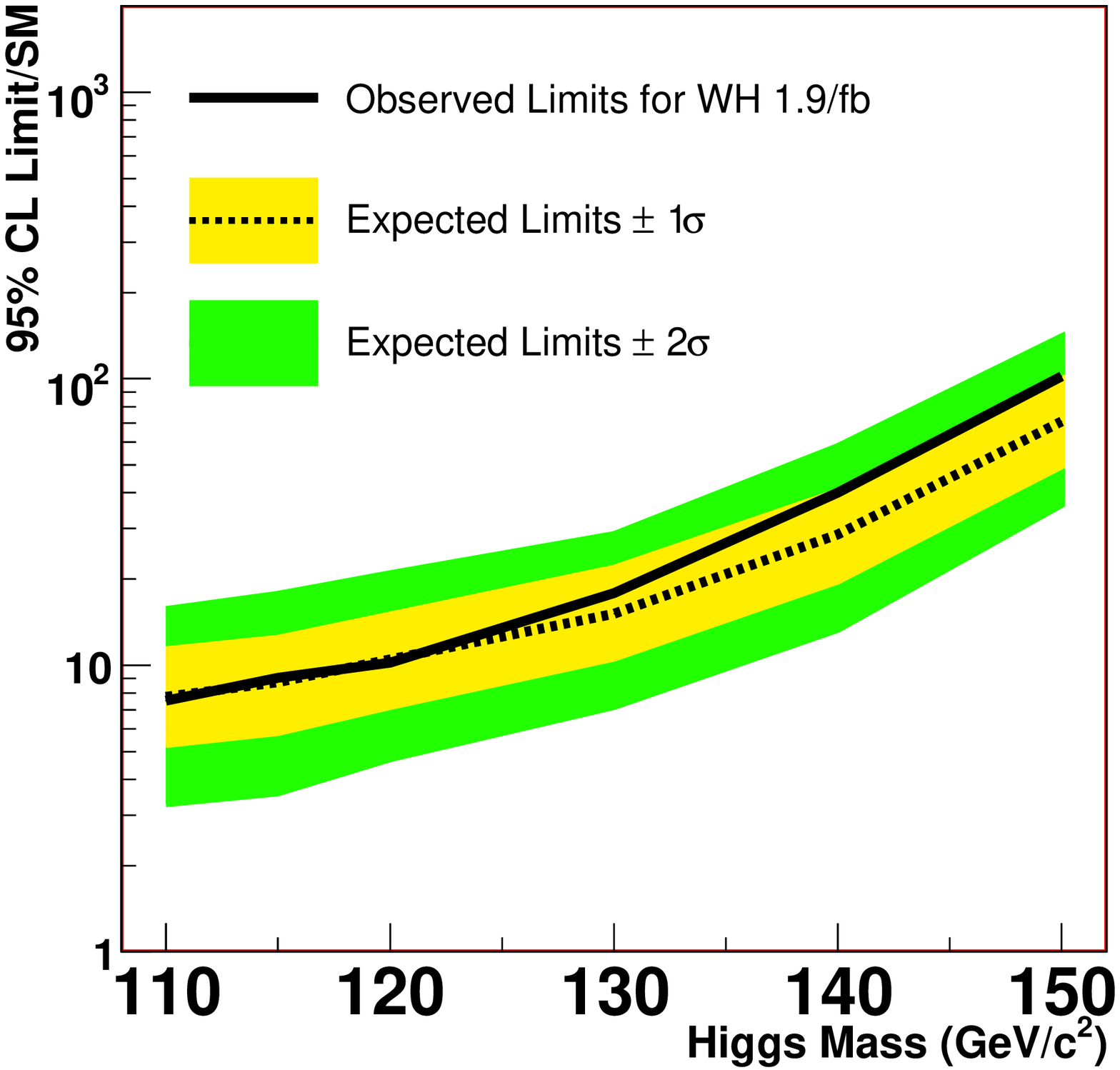}
   \caption{The observed (solid) and expected (dashed) 95\% confidence level upper limit on
     $\sigma(p\bar{p}\rightarrow WH)\cdot {\cal B}(H\rightarrow
     b\bar{b})$ relative to the standard model expectations with an integrated luminosity of $1.9$~fb$^{-1}$ along with 
the one and two $\sigma$ bands of the distributions of 
  expected outcomes from pseudoexperiments. 
     }
   \label{fig:upperLimit}
  \end{center}
 \end{figure}

The observed limits as a function of the Higgs boson
mass are shown in Fig.~\ref{fig:upperLimit} and
Table~\ref{tbl:upperLimit}, together with the expected limits
determined from pseudoexperiments.  We set 95\% confidence level upper limits on the
production cross section times branching fraction ranging from 1.2 to 1.1~pb or 
7.5 to 101.9 times the standard model expectation for Higgs boson masses from 
110 to $150\,\mathrm{GeV}/c^2$, respectively.

\begin{table}
  \begin{center}
   \caption{Observed and expected upper limits on $\sigma(p\bar{p}\rightarrow WH)\cdot {\cal B}(H\rightarrow b\bar{b})$ at 95~\%~C.L. compared to the standard model production rate calculated at NLO.}
    \begin{tabular}{cccccc}
      \hline
      \hline
      Higgs Mass    &     \multicolumn{2}{c}{Upper Limit (pb)} & \multicolumn{2}{c}{Upper Limit/SM}\\ \hline
       GeV/c$^{2}$  &      Observed  & Expected   & Observed & Expected \\
      \hline
      110  &               1.2       &  1.2   & 7.5 & 7.8 \\
      115  &               1.2       &  1.1   & 9.0 & 8.7 \\
      120  &               1.1       &  1.1   & 10.2 & 10.5  \\
      130  &               1.1       &  0.9   & 17.9 & 15.2  \\
      140  &               1.2       &  0.8   & 40.1 & 28.7  \\
      150  &               1.1       &  0.8   & 101.9 & 70.9  \\
      \hline \hline
    \end{tabular}
    \label{tbl:upperLimit}
  \end{center}
\end{table}

\section{Conclusions}
\label{sec:conclusions}
We have presented a search for the standard model Higgs boson in the
$\ell \nu b\bar b$ final state expected from $WH$ production at CDF. 
The search sensitivity is improved significantly with respect to
previous searches, by about 60\% more than the expectation from simple luminosity
scaling.  The main improvements are using jet probability $b$-tagging, a multivariate neural network technique to 
further enhance sensitivity to the signal, and increasing the acceptance for signal events by including leptons in the forward region of the detector. 
These improvements, along with a dataset
of $1.9\,\mathrm{fb}^{-1}$, allow us to 
set a 95\% confidence level upper
limit on the production cross section times branching fraction that
ranges from 1.2 to 1.1~pb or 7.5 to 101.9 times the standard model expectation
for Higgs boson masses spanning from 110 to
$150\,\mathrm{GeV}/c^2$, respectively.

This channel is an important component of the combined search for the Higgs boson at the Tevatron~\cite{Group:1}. With the Tevatron expected to deliver a factor of 4 more data, it is possible that the combined search with the complete dataset will be sensitive to the standard model Higgs cross section in this low 
mass range, which is of particular interest according to indications derived 
from standard model fits to electroweak observables~\cite{Alcaraz:2008mx}.

\begin{acknowledgments}
We thank the Fermilab staff and the technical staffs of the participating institutions for their vital contributions. This work was supported by the U.S. Department of Energy and National Science Foundation; the Italian Istituto Nazionale di Fisica Nucleare; the Ministry of Education, Culture, Sports, Science and Technology of Japan; the Natural Sciences and Engineering Research Council of Canada; the National Science Council of the Republic of China; the Swiss National Science Foundation; the A.P. Sloan Foundation; the Bundesministerium f\"ur Bildung und Forschung, Germany; the Korean Science and Engineering Foundation and the Korean Research Foundation; the Science and Technology Facilities Council and the Royal Society, UK; the Institut National de Physique Nucleaire et Physique des Particules/CNRS; the Russian Foundation for Basic Research; the Ministerio de Ciencia e Innovaci\'{o}n, and Programa Consolider-Ingenio 2010, Spain; the Slovak R\&D Agency; and the Academy of Finland. 
\end{acknowledgments}

\bibliography{reference}

\end{document}